\documentclass[preprint,eqsecnum,preprintnumbers,nofootinbib,byrevtex,prd,aps,showpacs,showkeys,groupedaddress,floatfix]{revtex4}
\usepackage{bm}
\usepackage{graphics}
\usepackage{graphicx}
\usepackage{epsfig}
\usepackage{amssymb}
\usepackage{amsmath}

\newcommand\nn{\nonumber}
\newcommand\ba{\begin{eqnarray}}
\newcommand\ea{\end{eqnarray}}

\begin{document}

\title{The inclusive production of charged pion pairs in proton-antiproton collisions}
\author{A.~I.~Ahmadov$^{1}$~\footnote{E-mail: ahmadovazar@yahoo.com}}
\author{C.~Aydin$^{2}$~\footnote{E-mail: coskun@ktu.edu.tr}}
\author{O.~Uzun$^{2}$~\footnote{E-mail:$\mbox{oguzhan}_-\mbox{deu@hotmail.com}$}}
\affiliation {$^{1}$ Department of Theoretical Physics, Baku State University, Z. Khalilov st. 23, AZ-1148, Baku, Azerbaijan\\
$^{2}$ Department of Physics, Karadeniz Technical University, 61080, Trabzon, Turkey}

\date{\today}

\begin{abstract}
In this study, we have considered the contribution of the
higher-twist (HT) effects of the subprocesses to inclusive pion pair
production cross section in the high energy proton-antiproton
collisions by using various pion distribution amplitudes (DAs)
within the frozen coupling constant approach and compared them with
the leading-twist contributions. The feature of the HT effects may
help the theoretical interpretation of the future PANDA experiment.
The dependencies of the HT contribution on the transverse momentum
$p_T$, the center of mass energy $\sqrt s$, and the variable $x_T$
are discussed numerically with special emphasis put on DAs.
Moreover, the obtained analytical and numerical results for the
differential cross section of the pion pair production are compared
with the elastic backward scattering of the pion on the proton. We
show that the main contribution to the inclusive cross section comes
from the HT direct production process via gluon-gluon fusion. Also,
it is strongly dependent on the pion DAs, momentum cut-off parameter
$\triangle p$ and $<{q_{T}^2}>$ which is the mean square of the
intrinsic momentum of either initial parton.
\end{abstract}

\pacs{12.38.-t, 13.60.Le, 14.40.Aq, 13.87.Fh}
\keywords{higher-twist, pion distribution amplitude, renormalization
scale} \maketitle

\section{\bf Introduction}

It is well - known that quantum chromodynamics (QCD) is the
fundamental theory of strong interactions. QCD describes the strong
interactions between quarks and gluons, also the structure and
dynamics of hadrons at the amplitude level.

The hadronic distribution amplitude (DA) in terms of internal
structure degrees of freedoms is important in QCD process
predictions. Parton DAs are important ingredients in applying QCD to
hard exclusive processes via the factorization theorem
\cite{Lepage11,Efremov1,Efremov2}. Understanding of the hadronic
structure in terms of the fundamental degrees of freedom of QCD is
one of the fascinating questions of the popular research area in
physics. The important processes of the perturbative quantum
chromodynamics (pQCD) are hadron pair production at large transverse
momenta in hadron-hadron collisions. While parton distributions at
leading-twist (LT) are basically relevant to the description to the
accuracy of leading power and refer to parton configurations with
the minimal number of constituents. However, the higher-twist (HT)
distributions are more numerous and they are used to consider the
various effects owing to parton virtuality, transverse momentum, and
contributions from higher Fock states which are relevant to describe
the power-suppressed corrections in the hard momentum. Braun
\textit{et al.} \cite{Braun1,Braun2,Braun3} recognized the important
role of the LT and the HT parton distributions in hard exclusive
process. The existing theoretical framework for the DA description
is based on the conformal symmetry of the QCD Lagrangian for an
exhaustive review
\cite{Lepage11,Efremov1,Efremov2,Brodsky11,Colangelo,Bauer,Feldman}.

The main difficulty in making precise perturbative QCD predictions
is the uncertainty in determining the renormalization scale $\mu$ of
the running coupling $\alpha_s(\mu^2)$. In practical calculations,
it is difficult to guess a simple physical scale of the order of a
typical momentum transfer in the process. Then we need to vary this
scale over a range $Q/2$, $2Q$. In a common case, this problem for
all orders was solved in Refs. \cite{Brodsky1,Mojaza}. Evolution
kernels are the main tools of the well-known evolution equations for
the parton distribution in deep inelastic scattering processes and
for the parton distribution amplitudes in hard exclusive reactions.
The Dokshitzer-Gribov-Lipatov-Altarelli-Parisi
(DGLAP)~\cite{Gribov,Lipatov,Dokshitzer,Altarelli} equation describe
the dependence of the parton distributions on the renormalization
scale $\mu^2$. Until now, DGLAP evolution equations have been known
as the most successful and major tools to study the structure
functions of hadrons and ultimately structure of matter,
ultra-high-energy cosmic rays. Also, the DGLAP  equations describe
the influence of the perturbative QCD corrections on the
distribution functions that enter the parton model of deep inelastic
scattering processes defined in the form as
\ba
\frac{d}{dln\mu^2}G_{i}(x,\mu^2)=\frac{\alpha_s(\mu^2)}{2\pi}\int_{x}^{1}\frac{dy}{y}G_{i}(y,\mu^2)P_{qq}(\frac{x}{y}) .
 \ea
From Eq. (1.1), we obtain an integrodifferential equation in the logarithm of the virtuality
\ba
\frac{d}{dln\mu^2}xG(x,\mu^2)=\frac{\alpha_s}{2\pi}\int_{x}^{1}dy\biggl[\sum_{f}P_{gq}(\frac{x}{y})\left(\frac{x}{y}q_{f}(\frac{x}{y},\mu^2)+\frac{x}{y}\bar{q}_{f}(\frac{x}{y},\mu^2)\right)+\nn \\
P_{gg}(\frac{x}{y})\frac{x}{y}G(\frac{x}{y},\mu^2)\biggr].
 \ea
Analogously, one finds for the quark and antiquark distributions as
\ba
\frac{d}{dln\mu^2}xq_{f}(x,\mu^2)=\frac{\alpha_s}{2\pi}\int_{x}^{1}dy\left[P_{qq}(\frac{x}{y})\frac{x}{y}q_{f}(\frac{x}{y},\mu^2)+P_{qg}(\frac{x}{y})\frac{x}{y}G(\frac{x}{y},\mu^2)\right],
 \ea
\ba
\frac{d}{dln\mu^2}x\bar{q}_{f}(x,\mu^2)=\frac{\alpha_s}{2\pi}\int_{x}^{1}dy\left[P_{qq}(\frac{x}{y})\frac{x}{y}\bar{q}_{f}(\frac{x}{y},\mu^2)+P_{qg}(\frac{x}{y})\frac{x}{y}G(\frac{x}{y},\mu^2)\right].
 \ea

Here $P_{qq}(\frac{x}{y})$  and $P_{qg}(\frac{x}{y})$ are known as
DGLAP splitting functions. These differential equations describe to
leading-logarithmic accuracy the change in the parton distribution
functions when changing $\mu^2$. They are a significant example of
what one calls evolution equations in quantum field theory. Solving
them results to the resummation of all the leading-order collinear
QCD corrections to deep inelastic scattering processes.
Equivalently, the DGLAP equations can be regarded as
renormalization-group equations, which renormalize the parton
densities with respect to the scale $\mu^2$. The DGLAP equation. It
allows one to explain the phenomenon of the scaling violation of the
proton structure function.

The dependence of the DA on the factorization scale $\mu^2_F$ is
handled by the Efremov-Radyushkin-Brodsky-Lepage (ERBL) evolution
equation ~\cite{Lepage11,Efremov1,Efremov2} which is defined the
following form
\ba \frac{\partial \Phi(x, \mu_{F}^2)}{\partial ln\mu_{F}^2}= \int
dy V(x,y,\alpha_s(\mu_{F}^2))\Phi(y,\mu_{F}^2). \ea

The evolution kernel $V(x,y,\alpha_s(\mu_{F}^2))$ is calculable in
perturbation theory \ba
V(x,y,\alpha_s(\mu_{F}^2))=\frac{\alpha_s(\mu_{F}^2)}{\pi}V_{1}(x,y)+
\left(\frac{\alpha_s(\mu_{F}^2)}{\pi}\right)^2V_{2}(x,y)\ea The
one-loop evolution kernel $V_0$ was introduced in
Refs.~\cite{Lepage11,Brodsky11}, an analogous expression for $V_2$
at the two-loop level was derived in
Refs.~\cite{Dittes,Sarmadi,Katz,Radyushkin}. It should be noted that
the  HT refers to contributions suppressed by powers of large
momentum with respect to the leading-twist. The leading-twist (LT)
is a standard processes of the pQCD within the collinear
factorization where hadrons are produced through fragmentation
processes. However, HT processes are taken usually as direct hadron
production, where the hadron is produced directly in the hard
subprocess rather than by quark/gluon fragmentation. Higher-twist
dynamics at the hadron production in hadron-hadron collisions is
widely studied in Ref.~\cite{Arleo}.

In  Refs. ~\cite{Collins,Boer}, it is showed that hard-scattering
factorization is disrupted in the production of high-$p_T$ hadrons
in the case of the hadrons being back-to-back by using $k_T$
factorization. It is worthy noted that perturbative QCD
factorization formulas are modified at leading twist by initial and
final state corrections. The explicit counterexample was provided
for the single-spin asymmetry with one beam transversely polarized
as well.

The calculation and analysis the contribution of the HT effects to
cross section on the dependence of the pion DA in inclusive pion
pair production at $p \bar{p}$ collision within the frozen coupling
constant (FCC) approach are important and interesting research
problems. Therefore, HT effects in QCD have been predicted and
computed in the last 40 years by many researchers for various
phenomenas~\cite{Bagger,Bagger1,Baier,Gupta,Sadykhov,Ahmadov1,Ahmadov2,Ahmadov3,Ahmadov4,Ahmadov5,Ahmadov6,Ahmadov7,Ahmadov8,Ahmadov9,Ahmadov10,Demirci}.
Meson pair production in photon-photon, nucleon-nucleon, and
proton-antiproton collisions have been studied from high to low
energies during the last few years, applying different approaches
such as HT mechanism, central exclusive production mechanism,
effective meson theory, and standard
pQCD~\cite{Ji,Buran,Wang,shyam,Bystritskiy1,Wang1,Khoze, Djagouri}.

Precision experimental studies of meson pair production in
proton-antiproton collisions at low energies are proposed in the
experiment named  PANDA \cite{PANDA}. The PANDA scientific program
use 1.5 - 15 GeV energy range for interactions between protons and
antiprotons where this energy lies near the pion production
threshold. This program include several measurements and it
addresses fundamental questions of QCD by obtaining the detailed
analysies of all possible mechanisms of meson pair production
\cite{Poslavsky}. In this study, we examine the contribution of the
HT effects to inclusive charged pion pair production at
proton-antiproton collisions by using different pion DAs obtained
within holographic and perturbative QCD which can be helpful for an
explanation of the PANDA experiment. We have also given theoretical
predictions of the inclusive charged pion pair production in
$p\bar{p}$ collisions by accounting for the leading order diagrams
in partonic cross sections.

The physical information of the inclusive pion pair production can
be obtained efficiently in the pQCD and it is, hence, possible to
compare directly with the experimental data. The corresponding
hard-scattering subprocesses occur via three different mechanisms.
The first one is the direct production of charged pion pairs which
are produced directly at the hard-scattering subprocess (see
Fig.\ref{fig:fig1}). The second one is the semidirect production of
charged pion pairs in which one pion is produced from jet
fragmentation (see Fig. \ref{fig:fig2}). Finally, the last one is
the double jet production and fragmentation where both pions are
produced from fragmentation of the final quarks or gluons. The first
two mechanisms are HT contributions and the last is the LT
contribution. Therefore, we must systematically compare these
different mechanisms. We use the frozen coupling constant (FCC)
approach during numerical evaluation in all calculations. In order
to obtain an accurate value of the ratio (HT/LT), we need to use the
fact that prompt pions appear "non-accompanied" by any other hadron,
while this is not valid for the general case in which particles are
resulting from the jet fragmentation. That criterion of
"non-accompaniment" into the general formalize a momentum cut-off
parameter $\triangle p$ is considered in calculation \cite{Engels}.

The rest of the paper is organized as follows: In Sec.\ref{ir}, a
brief review for the formalism used for the calculation of the HT
contribution to cross section and some formulas for the HT cross
section of the process $p\bar {p}\to\pi^{+}\pi^{-}X$ is given. In
Sec. ~\ref{ht}, some formulas for LT  cross sections for pion pairs
production are provided. In Sec. \ref{com}, we present a comparison
of the HT charged pion pair production $p\bar{p}\to\pi^{+}\pi^{-}$
cross section with elastic $\pi^{\pm} p \to \pi^{\pm}p$ cross
section, and the numerical results for the cross section and the
discussion of the dependence on the cross section on the pion DA are
provided in Sec.~\ref{results}. Finally, the concluding remarks are
stated in Sec.~\ref{conc}.

\section{Higher-Twist  Contribution to Inclusive Direct  Pion Pair  Production Cross Section}
\label{ir}

The inclusive production of charged pion pair with large transverse
momenta $(p_T >1$ \,\,GeV/$\emph{c})$ in opposite hemispheres,
essentially back-to-back in the center - of - mass system of the
incoming hadrons is considered in this study. This mechanism had
been already analyzed in \cite{Engels} for the case of the two
particle back-to-back cross - section reflecting the
$p_T$-dependence of the hard scattering subprocesses undisturbed by
the internal momenta of the constituents. There are many other
studies in the literature about physical properties of FCC
~\cite{Curci,Curci1,Greco1,Greco2,Webber1,Webber2,Mandelstam,Brown,Brown1,Cornwall,Alkofer,
Smekal,Parisi,Schwinger}. In numerically, calculating the HT cross
section (within FCC approach for the square of the transfer momentum
of the hard gluon) and LT cross section we can use the following
values as
\ba
Q^2=\left\{\begin{array}{ll}\frac{1}{2}p^{2}_T \, , & \,\,\,{\mbox {for direct HT contribution}} \\
\frac{1}{2}\frac{p^{2}_T}{\sqrt z} \, , &
\,\,\,{\mbox {for semi-direct HT contribution}} \\
\frac{1}{2}\frac{p^{2}_T}{\sqrt {zz'}}, \,& \,\,\ {\mbox {for LT contribution.}}
\end{array}
\right. \ea

Using the fact that prompt pions are non-accompanied by any other
hadron, the ratio contributions of HT and LT can be calculated
accurately. However, this is not valid for the general case in which
particles are occurring from the jets fragmentation. This criterion
can be incorporated into the general formulas via a momentum cut-off
parameter $\triangle p$ [29]. The details of analytical calculations
on HT and LT contributions will be given in the following
subsections. The leading order HT Feynman subdiagrams for the
inclusive direct pion pair
\begin{figure}[!hbt]
    \begin{center}
\includegraphics[scale=0.55]{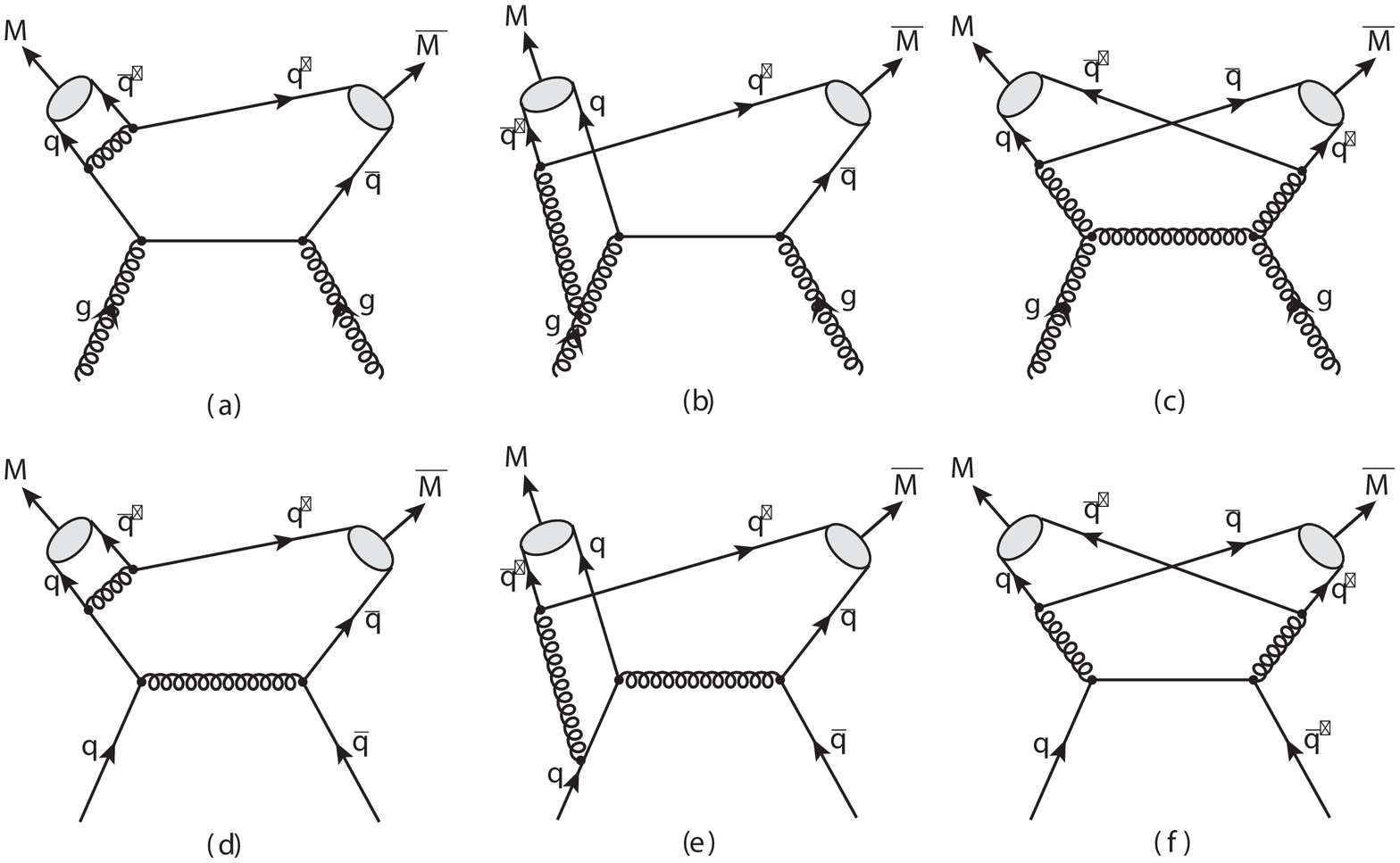}
     \end{center}
\caption{QCD Feynman diagrams  of the partonic process
$\text{g}\text{g} \to M\bar M$ and $q\overline q \to M\bar M$ for
direct meson pair production at leading order.}\label{fig:fig1}
\end{figure}
\begin{figure}[!hbt]
    \begin{center}
\includegraphics[scale=0.55]{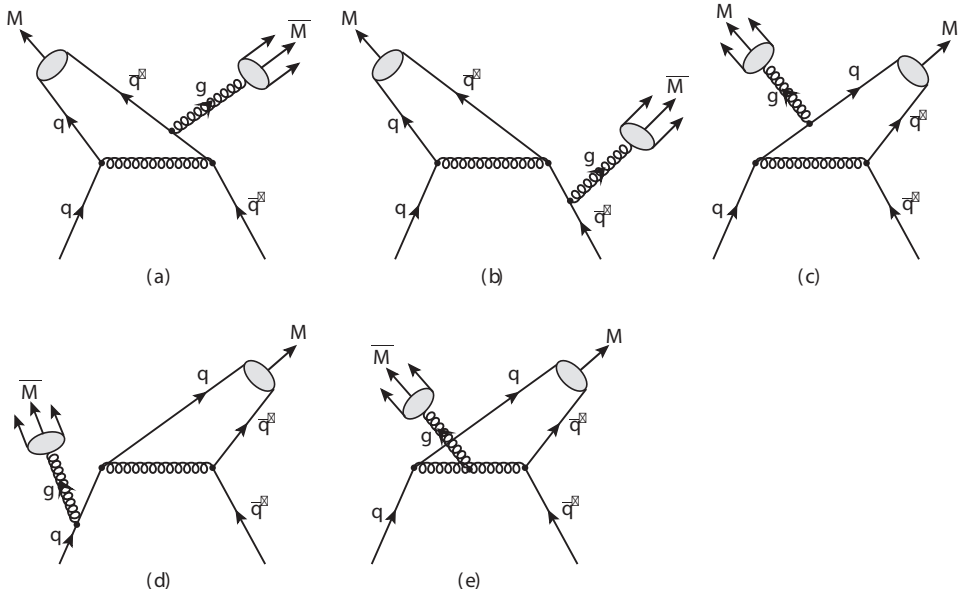}
     \end{center}
\caption{QCD Feynman diagrams  of the partonic process $q\bar q\to
M\bar M$ for semi-direct meson pair production at leading
level.}\label{fig:fig2}
\end{figure}
production in the proton-antiproton collision $p\bar p
\to\pi^{+}\pi^{-}X$ are taken as $gg\to \pi^+\pi^-$  and
$q\bar{q}\to\pi^+\pi^-$ (where $q$ is either $u$ or $d$ quarks)
which contributes to the main process (see Fig. \ref{fig:fig1}).
Semi-direct pion pair productions in the same process are shown in
Fig. \ref{fig:fig2}. The amplitude for this subprocess can be
obtained by using the Brodsky-Lepage formula~\cite{Brodsky11}
\begin{equation}
M(\hat s,\hat
t)=\int_{0}^{1}{dx_1}\int_{0}^{1}dx_2\delta(1-x_1-x_2)\Phi_{M}(x_1,x_2,Q^2)T_{H}(x_1,x_2;Q^2,\mu_{R}^2,\mu_{F}^2),
\end{equation}
where $T_H$ is the sum of the graphs contributing to the
hard-scattering part of the subprocess. At the leading order of pQCD
calculations, the hard scattering amplitude $T_{H}(x_1,x_2;Q^2,
\mu_{R}^2,\mu_{F}^2)$ does not depend on the factorization scale
$\mu_{F}^2$, but strongly depends on $\mu_{R}^2$. However, the
scales $\mu_{F}^2$ and $\mu_{R}^2$ are independent of each other.

In principle, all measurable quantities in QCD should be invariant
under any choice of renormalization scale and scheme. It is clear
that the use of different scales and schemes may lead to different
theoretical predictions. Therefore, the constructive mathematical
tool for defining QCD is a choice of the renormalization scale which
makes scheme independent results at all fixed order in running
coupling constant $\alpha_{s}$. For direct pion pair production, the
subprocesses are taken as $gg \to \pi^+\pi^-$, $u\bar{u}
\to\pi^+\pi^-$ and $d\bar{d} \to \pi^+\pi^-$. However, for the
semi-direct pion pair production the subprocesses are $q \bar{q} \to
\pi g$, $qg \to \pi q^{'}$ and $\bar {q}g \to \pi \bar q^{'}$. In
the processes $q \bar{q} \to \pi g$, the final gluon is $qg \to \pi
q^{'}$, the final quark is $\bar {q}g \to \pi \bar q^{'}$, and the
final antiquark is taken as a fragmentation of the pion. Here, $q,
\bar{q}$ and $g$ are the constituent of the initial target proton
and anti-proton. It should be noted that, each $q\bar q$ pair is
collinear and has the appropriate color, spin, and flavor content
projected out to form the parent pion. The production of the pair of
pion or jets in the large transverse momentum is available at the
high energy, especially at the CERN Large Hadron Collider. In the
direct pion pair production case, the hadronic pion is the final
product of the hard-scattering processes. But in the final state of
the semi-direct pion pair production, one of the hadronic gluon or
jets are fragmented to a pion. Dynamical properties of the jet are
close to the parent parton which are carried by one of  part of the
four-momentum of the parent parton. In order to explain parton level
kinematics, we use the pion pair production process considered in
\cite{Owens}.

The parton-level differential cross sections for the direct pion
pair production are obtained as
\ba
\frac{d\sigma}{dcos\theta}(gg\to\pi^+\pi^-) =  \nn \\
\frac{256\pi^3\alpha_{s}^4f_{\pi}^4}{23328}\biggl[\int_{0}^{1}\frac{\Phi_{\pi}(x,Q^2)
dx}{x(1-x)}\biggr]^2\biggl[\int_{0}^{1}dx\int_{0}^{1}dy\frac{\Phi_{\pi}(x,Q^2)\Phi_{\pi}(y,Q^2)}{x(1-x)y(1-y)}\cdot\frac{x(1-x)+y(1-y)}{xy+(1-x)(1-y)}\biggr]^2
, \,\,\,\,\,\, \ea
\ba
\frac{d\sigma}{dcos\theta}(q\overline q \to \pi^+\pi^-) =  \nn \\
 \frac{256\pi^3\alpha_{s}^4f_{\pi}^4}{139968}\biggl[\int_{0}^{1}\frac{\Phi_{\pi}(x,Q^2)
dx}{x(1-x)}\biggr]^2\biggl[\int_{0}^{1}dx\int_{0}^{1}dy\frac{\Phi_{\pi}(x,Q^2)\Phi_{\pi}(y,Q^2)}{x(1-x)y(1-y)}\cdot\frac{x(1-x)+y(1-y)}{xy+(1-x)(1-y)}\biggr]^2  \nn \\
\cdot [7-16xy-\frac{1}{xy+(1-x)(1-y)}[2x(1-2y(x+y))-4x^2+4xy]].
\,\,\,\,\, \ea
Similarly, for the semi-direct  pion pair production case which
corresponds to the Feynman diagrams in Fig. \ref{fig:fig2}, the hard
collisions subprocesses are taken in three different ways as,
\begin{enumerate}

\item $q\bar{q'}\to \pi^+(\pi^-)g$, where the gluon is fragmented
to a pion $(g\to\pi^-(\pi^+))$,

\item $qg\to\pi^{\pm}q'$, where  quark
is fragmented to pion $(q'\to\pi^{\mp})$,

\item $\bar{q}g\to\pi^{\pm}\bar q'$, $(\bar q'\to\pi^{\mp})$, where the antiquark is
fragmented to a pion.
\end{enumerate}
The corresponding differential cross sections of the subprocesses
are defined for these cases as
\ba \frac{d\sigma}{dcos\theta}(q\overline{q'}\to \pi^\pm
g)=\frac{128\pi^2\alpha_{s}^3 f_{\pi}^2}{729{\hat
s}^2}\left[\int_{0}^{1}\frac{\Phi_{\pi}(x,Q^2) dx}{x(1-x)}\right]^2,
\ea
\ba \frac{d\sigma}{dcos\theta}(qg\to \pi^\pm
q')=\frac{80\pi^2\alpha_{s}^3 f_{\pi}^2}{3888{\hat
s}^2}\left[\int_{0}^{1}\frac{\Phi_{\pi}(x,Q^2) dx}{x(1-x)}\right]^2,
\ea
\ba \frac{d\sigma}{dcos\theta}(\bar{q} g\to \pi^\pm \bar
q')=\frac{80\pi^2\alpha_{s}^3 f_{\pi}^2}{3888{\hat
s}^2}\left[\int_{0}^{1}\frac{\Phi_{\pi}(x,Q^2) dx}{x(1-x)}\right]^2,
\ea respectively. The main goals of this study are the calculation
and also, if possible, extraction of the contributions HT effects to
the cross section by the FCC approach using different pion DAs. For
the calculation of the cross section,
 we need to apply the factorization formula which was predicted by Gunion and Petersson ~\cite{Petersson,Carimalo}.
 In this approach a differential cross section of the process $p \bar p \to \pi^{+}\pi^{-} X$ is defined as
\ba \Sigma_{\pi^+\pi^-}=E_CE_D\frac{d\sigma}{d^3p_Cd^3p_D}= \nn \\
=\frac{1}{\pi^2s<{q^{2}_T}>}\int_{z_{min}}^{1}\frac{dz}{z^2}\int_{z_{min}}^{1}\frac{dz'}{z'^2}
F(z,z')G_{{q_{1}}/{p_{1}}}(x_{1},Q^2) G_{{q_{2}}/{p_{2}}}(x_{2},Q^2) \times \nn \\
\times \frac{d\sigma}{dcos\theta}(q\bar{q}(gg) \to
\pi^+\pi^-)D_{M/C}(z,Q^2)D_{\bar{M}/D}(z',Q^2),\,\, \ea
where $s$ is the center-of-mass energy squared of main process,
$<{q_{T}^2}>$ is the mean square of the intrinsic momentum of either
initial parton $q_1, q_2$, $G_{q_1/p_1}$ and $G_{q_2/p_2}$ are the
universal PDFs for the partons $q_1$, $q_2$ in the proton and
antiproton $p_1$, $p_2$, respectively. They depend on the
longitudinal momentum fractions of the two partons in the case when
final jets are fragmenting to pion pair $x_1= x_2=2p_T/\sqrt{zz's}$
and on the scale parameter $Q^2$ of the central collision process.
$d\sigma/dcos\theta$ is the differential cross section of the
process and $\theta$ is the scattering angle. In the main process,
both pions are emitted  at $90^\circ$ in the center-of-mass  frame.
For the dependence of the symmetric pair production cross section
$E_{C}E_{D}\frac{d\sigma}{d^3p_{C}d^3p_{D}}$ at $90^\circ$ of the
transverse momentum, we take into account
$p_T=p_{T_{C}}=-p_{T_{D}}$, $y_C=y_D=0$, $\varphi_C=0$ and
$\varphi_D=\pi$.

The longitudinal momentum fractions of partons are defined in this form:
\ba
 x_1=-\frac{1}{2}(x_{T_1}e^{y_1}+x_{T_2}e^{y_2}),\\
x_2=-\frac{1}{2}(x_{T_1}e^{-y_1}+x_{T_2}e^{-y_2}), \ea in which
$y_1$, $y_2$  are the rapidities of the final particles.

For the calculation of the HT cross sections in the case of direct
pion pair production, we assume in Eq.(2.8) that $M=\pi^+$,
$C=\pi^+$ and $\bar {M}=\pi^-$, $D=\pi^-$. Therefore instead of
fragmentation functions (FFs) $D_{M/C}(z,Q^2)$ and
$D_{\bar{M}/D}(z',Q^2)$, we make the substitutions
$D_{\pi^+/\pi^+}(z,Q^2)=\delta(1-z)$ and
$D_{\pi^-/\pi^-}(z',Q^2)=\delta(1-z')$. But, for the HT cross
section in the semi-direct pion pair production case, we take
$M=\pi^+$, $C=\pi^+$, then we make the substitutions
$D_{\pi^+/\pi^+}(z,Q^2)=\delta(1-z)$. In the numerical calculations,
the function fragmentation of the gluon and quark ~\cite{Kniehl}
into a pion have been used. The function $F(z,z')$ called as the
correlation function is defined as
\ba
F(z,z')=\frac{z+z'}{2\sqrt{zz'}}\exp\left[\frac{-(z-z')^2p_{T}^2}{2z^2z'^2<q_{T}^2>}\right].
\ea
In the LT subprocess, the pion is indirectly emitted from the quark with fractional momentum $z$.
 The minimum value of the momentum fraction of the final parton $z_{min}$ is defined in this form:
\ba z_{min}=\frac{p_T}{p_T+\triangle p}. \ea
\noindent here $\triangle p$ is a momentum cut-off parameter which
describes the experimental upper limit for non-detection of one or
more particles accompanying either pion detected. It is assumed that
whenever this limit is exceeded, the corresponding event will be
rejected.

\section{LEADING-TWIST CONTRIBUTIONS TO INCLUSIVE CHARGED PION PAIR PRODUCTION CROSS SECTION}
\label{ht}

It is an important task to compare the HT corrections with LT contributions and to extract the HT corrections to the pion pair production cross section.

For the LT  cross section for the production of pion pairs, we take
the next subprocesses in which the final particles are fragmented to
pion pairs as $q\bar{q}\to gg$ ($g\to \pi^+$, $g \to \pi^-$), $gg
\to q\bar{q}$ ($q\to \pi^+$, $\bar q \to \pi^-$), $qg\to qg$ ($q\to
\pi^+$, $g \to \pi^-$), $gg \to gg$ ($g\to \pi^+$, $g \to \pi^-$)
and $q\bar{q}\to q\bar{q}$ ($q\to \pi^+$, $\bar q \to \pi^-$).

The corresponding differential cross section of the LT subprocesses
are written as ~\cite{Owens}
\ba \frac{d\sigma}{dcos\theta}(q_1q_2 \to
q_1q_2)=\frac{2\pi\alpha_{s}^2}{{9\hat
s}}\left(\frac{u^2+s^2}{t^2}\right), \ea
\ba \frac{d\sigma}{dcos\theta}(q_1\overline{q}_2 \to
q_1\overline{q}_2)=\frac{2\pi\alpha_{s}^2}{{9\hat
s}}\left(\frac{u^2+s^2}{t^2}\right), \ea
\ba \frac{d\sigma}{dcos\theta}(q_1q_1 \to
q_1q_1)=\frac{\pi\alpha_{s}^2}{2{\hat
s}}\left(\frac{4}{9}\cdot\left(\frac{u^2+s^2}{t^2}+\frac{s^2+t^2}{u^2}\right)-\frac{8}{27}\cdot\frac{s^2}{ut}\right),
\ea
\ba \frac{d\sigma}{dcos\theta}(q_1\overline{q}_1 \to
q_2\overline{q}_2)=\frac{2\pi\alpha_{s}^2}{{9\hat
s}}\left(\frac{u^2+t^2}{s^2}\right), \ea
\ba \frac{d\sigma}{dcos\theta}(q_1\overline{q}_1 \to
q_1\overline{q}_1)=\frac{\pi\alpha_{s}^2}{2{\hat
s}}\left(\frac{4}{9}\cdot\left(\frac{u^2+s^2}{t^2}+\frac{u^2+t^2}{s^2}\right)-\frac{8}{27}\cdot\frac{u^2}{st}\right),
\ea
\ba \frac{d\sigma}{dcos\theta}(q\overline{q} \to
gg)=\frac{\pi\alpha_{s}^2}{2{\hat
s}}\left(\frac{32}{27}\cdot\frac{u^2+t^2}{ut}-\frac{8}{3}\cdot\frac{u^2+t^2}{s^2}\right),
\ea
\ba \frac{d\sigma}{dcos\theta}(gg \to
q\overline{q})=\frac{\pi\alpha_{s}^2}{2{\hat
s}}\left(\frac{1}{6}\cdot\frac{u^2+t^2}{ut}-\frac{3}{8}\cdot\frac{u^2+t^2}{s^2}\right),
\ea
\ba
\frac{d\sigma}{dcos\theta}(qg\to qg)=\frac{\pi\alpha_{s}^2}{2{\hat
s}}\left(-\frac{4}{9}\cdot\frac{u^2+s^2}{us}+\frac{u^2+s^2}{t^2}\right),
\ea
\ba \frac{d\sigma}{dcos\theta}(gg\to
gg)=\frac{\pi\alpha_{s}^2}{{\hat
s}}\frac{9}{4}\left(3-\frac{ut}{s^2}-\frac{us}{t^2}-\frac{st}{u^2}\right),
\ea
\ba \frac{d\sigma}{dcos\theta}(q_1q_1\to
q_2q_2)=\frac{2\pi\alpha_{s}^2}{9{\hat
s}}\left(\frac{u^2+t^2}{s^2}\right). \ea
where subscripts 1 and 2 denote distinct flavors. The initial and
final state colors and spins have been averaged and summed,
respectively. Over the last few years, a great deal of progress has
been made in the investigation of the properties of hadronic wave
functions. The notion of distribution amplitudes refers to momentum
fraction distributions of partons in the meson, in particular, the
Fock state with a fixed number of components. For the minimal number
of constituents, the distribution amplitude $\Phi$ is related to the
Bethe-Salpeter wave function $\Phi_{BS}$ by

\begin {equation}
\Phi(x)\sim\int^{|k_{\perp}|<{\mu}}d^{2}k_{\perp}\Phi_{BS}(x,k_{\perp}).
\end {equation}

The standard approach to distribution amplitudes, which is due to
Brodsky and Lepage~\cite{Brodsky4}, considers the hadron's parton
decomposition in the infinite momentum frame. A conceptually
different, but mathematically equivalent formalism is the light-cone
quantization~\cite{Brodsky5}. The meson distribution amplitudes play
a key role in the hard-scattering QCD processes because they
encapsulate the essential nonperturbative features of the meson's
internal structure in terms of the parton's longitudinal momentum
fractions $x_i$. Meson DAs have been extensively studied by using
QCD sum rules. The original suggestion by Chernyak and Zhitnitsky of
a "double-humped" wave function of the pion at a low scale, far from
the asymptotic form, was based on an extraction of the first few
moments from a standard QCD sum rule approach~\cite{chernyak}, in
the Bakulev-Mikhailov-Stefanis(BMS) DA two non-trivial Gegenbauer
coefficients $a_2$ and $a_4$ have been extracted from the CLEO data
on the $\gamma\gamma^{\star} \to \pi^0$ transition form factor in
which the authors have used the QCD light-cone sum rules approach
and have included in their analysis the next to leading order
perturbative and twist-four corrections. Thus, in our numerical
calculations, we used several choices, such as the asymptotic DAs
predicted by pQCD evaluation, light-cone formalism, the light-front
quark model \cite{Lepage11}, the
Vega-Schmidt-Branz-Gutsche-Lyubovitskij (VSBGL) DA ~\cite{ Vega},
holographic meson DAs is obtained in the context of AdS/CFT ideas
~\cite{Brodsky2,Brodsky3} are studied considering two kinds of
holographic soft-wall models, the
Chernyak-Zhitnitsky(CZ)~\cite{chernyak}, and the BMS
~\cite{Bakulev}:
\ba
\Phi_{asy}(x)=\sqrt{3}f_{\pi}x(1-x),
\label{asy}
\ea
\ba
\Phi_{VSBGL}^{hol}(x)=\frac{A_1k_1}{2\pi}\sqrt{x(1-x)}exp\left(-\frac{m^2}{2k_{1}^2x(1-x)}\right),
\ea
\ba
\Phi^{hol}(x)=\frac{4}{\sqrt{3}\pi}f_{\pi}\sqrt{x(1-x)},
\ea
\ba
\Phi_{CZ}(x,\mu_{0}^2)=\Phi_{asy}(x)\left[C_{0}^{3/2}(2x-1)+\frac{2}{3}C_{2}^{3/2}(2x-1)\right],
\ea
\ba
\Phi_{BMS}(x,\mu_{0}^2)=\Phi_{asy}(x)\left[C_{0}^{3/2}(2x-1)+0.20C_{2}^{3/2}(2x-1)-0.14C_{4}^{3/2}(2x-1)\right].
\label{BMS} \ea

The pion DA can be expanded  over the eigenfunctions of the one-loop
ERBL equation
\ba
\Phi_{\pi}(x,Q^2)=\Phi_{asy}(x)\left[1+\sum_{n=2,4..}^{\infty}a_{n}(Q^2)C_{n}^{3/2}(2x-1)\right].
\ea
The evolution of the DA on the factorization scale $Q^2$ is handled
by the functions $a_n(Q^2)$ as
\begin {equation}
a_n(Q^2)=a_n(\mu_{0}^2)\left[\frac{\alpha_{s}(Q^2)}{\alpha_{s}(\mu_{0}^2)}\right]^{\gamma_n/\beta_0},
\end{equation}
$$
\frac{\gamma_2}{\beta_{0}}=\frac{50}{81},\,\,\,\frac{\gamma_4}{\beta_{0}}=\frac{364}{405},\,\,
n_f=3.
$$

In Eq.(3.18), $\gamma_n$'s are anomalous dimensions defined by the expression
\ba \gamma_n=C_F\left[1-\frac{2}{(n+1)(n+2)}+4\sum_{j=2}^{n+1}
\frac{1}{j}\right].\ea

The Gegenbauer moments  $a_n$  can be determined by using the
Gegenbauer polynomials orthogonality condition \ba \label{eq:ort}
\int^1_{-1}(1-\zeta^2)
C_{n}^{3/2}(\zeta)C_{n'}^{3/2}(\zeta)d\zeta=\frac{\Gamma(n+3)\delta_{nn'}}{n!(n+3/2)}.
\ea The Gegenbauer moments $a_n$ are very practical in studying the
DAs because they form the shape of the corresponding hadron wave
function. Hereby, it can be possible to derive from theoretical
models or extracted from the experimental data. Besides, these
moments reveal, how much the DAs deviate from the asymptotic one.
The strong coupling constant $\alpha_{s}(Q^2)$ at the one-loop
approximation is given as
\begin{equation}
\alpha_{s}(Q^2)=\frac{4\pi}{\beta_0{ln}(\frac{Q^2}{\Lambda^2})}.
\end{equation}
where $\Lambda$ is the QCD scale parameter, $\beta_0$ is the QCD
beta function one-loop coefficients. It should be noted that the
choice of renormalization scale in $\alpha_{s}(Q^2)$ is one of the
main problems in QCD. In the numerical calculations, the hard gluon
square momentum was used from Eq. (2.1). Notice that the pion DAs
presented in Eqs. (\ref{asy})-(\ref{BMS}) constructed from theory
and experiment strongly depend on the applied methods. However, the
correct pion wave function is still an open problem in QCD.

\section{COMPARISON HIGHER-TWIST PION PAIR PRODUCTION CROSS SECTION $p \bar p \to \pi^{+}\pi^{-}$ WITH ELASTIC  $\pi^{\pm} p \to \pi^{\pm}p$ CROSS SECTION}\label{com}

It would be important and interesting to compare the
proton-antiproton annihilation process $p \bar p \to \pi^{+}\pi^{-}$
with the elastic backward scattering $\pi^{\pm} p \to \pi^{\pm}p$
process by fixing $u$ and switching $s$ and $t$. In order to compare
matrix elements at given values $s$, $t$ or $u$, spin and
phase-space factors have to be taken into account more specifically.
We compare the differential cross section for the annihilation
process $\frac{d\sigma}{dt}(p \bar p \to \pi^{+}\pi^{-})$ with the
corresponding elastic backward cross section
$\frac{d\sigma}{dt}(\pi^{\pm}p \to \pi^{\pm}p)$ using the suitable
spin and phase-space factors. So,
\ba \frac{d\sigma}{dt}(p\bar p \to \pi^{+}\pi^{-}) =
\frac{(2s_{\pi}+1)(2s_{p}+1)}{(2s_{\bar
p}+1)(2s_{p}+1)}\left(\frac{k_{\pi p}}{p_{\bar p
p}}\right)^2\frac{d\sigma}{dt}(\pi^{\pm}p \to \pi^{\pm}p).\ea
where $s_{\pi}$ and $s_{p}$ ($s_{\bar p}$) are the spins of the pion
and proton(antiproton), $k_{\pi p}$ and $p_{\bar p p}$ are the
center of mass momenta, evaluated at the same center-of-mass energy.
If the hadrons are produced at $90^\circ$ with rapidities
$y_C=y_D=0$, the hard scattering cross section $d\sigma/d\hat t$ is
probed at angles around $90^\circ$ where $\hat t=\hat u =-\hat s/2$.
The comparison is relevant only at the center of mass energies and
therefore the elastic backward cross sections are scaled by using an
$s^{-2}$ dependence for $\pi p$. The result of the comparison are
present in Figs. \ref{fig:fig13} and \ref{fig:fig14}.

\section{NUMERICAL RESULTS AND DISCUSSION} \label{results}

\begin{figure}[!hbt]
\includegraphics[width=8.6 cm]{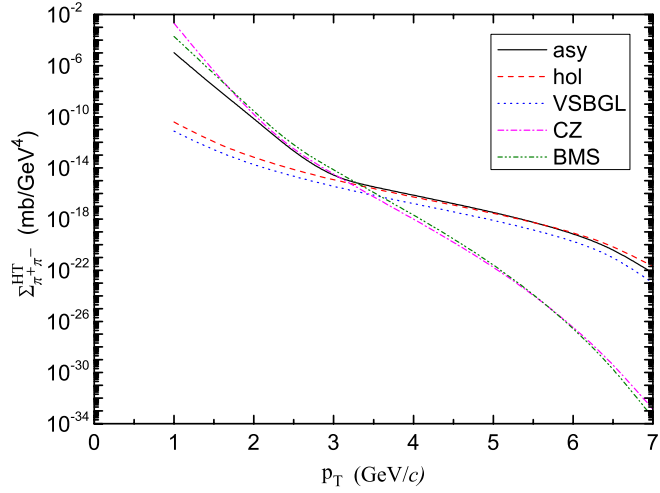}
\caption{HT contribution to charged pion pair production $p\bar
{p}\to\pi^{+}\pi^{-}X$  cross section $\Sigma_{\pi^+\pi^-}^{HT}$ as
a function of the transverse momentum $p_{T}$ for momentum cut-off
parameter $\triangle p=0.3$ GeV/\emph{c}, at $\sqrt s=15$\,\, GeV
and $y=0$} \label{fig:fig3}
\end{figure}

\begin{figure}[h]
\includegraphics[width=8.6 cm]{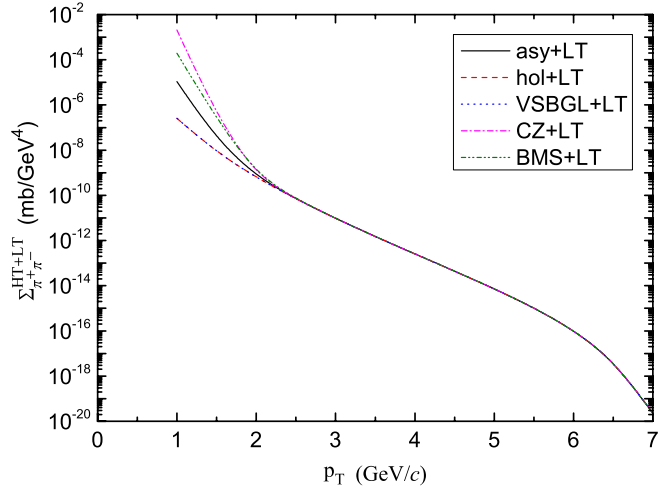}
\caption{The sum of HT and LT  contribution to charged pion pair
production $p\bar {p}\to\pi^{+}\pi^{-}X$  cross section
$\Sigma_{\pi^+\pi^-}^{HT+LT}$ as a function of the transverse
momentum $p_{T}$ for momentum cut-off parameter $\triangle
p=0.3$GeV/\emph{c}, at $\sqrt s$=15\,\, GeV and $y=0$. Notice that
curves for asy, hol, VSBGL, CZ and BMS pion DA in the region
$2\,\,$GeV/\emph{c}$<p_T<7\,\,$GeV/\emph{c} completely overlap.}
\label{fig:fig4}
\end{figure}

\begin{figure}[!hbt]
\includegraphics[width=8.6 cm]{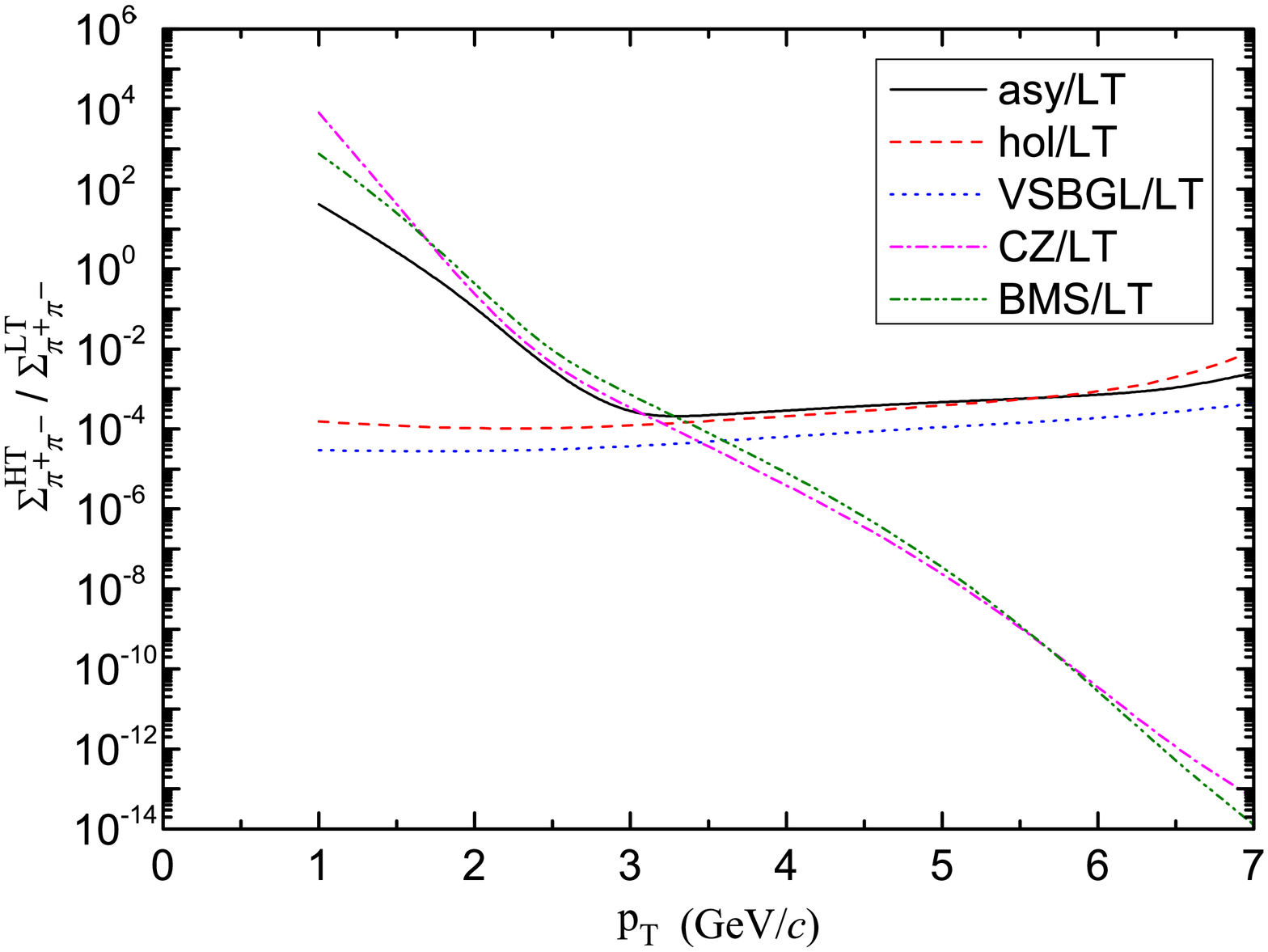}
\caption{Ratio $\Sigma_{\pi^+\pi^-}^{HT}/\Sigma_{\pi^+\pi^-}^{LT}$
as a function of the transverse momentum $p_{T}$ of the pion pair
for $<{q_{T}^2}>$=0.25 GeV$^2$/\emph{c}$^2$, at the c.m. energy
$\sqrt s$=15 GeV and $y=0$.} \label{fig:fig5}
\end{figure}

\begin{figure}[!hbt]
\includegraphics[width=8.6 cm]{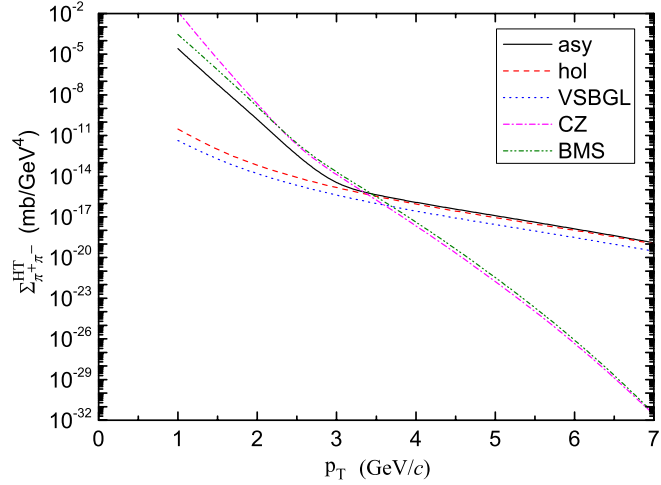}
\caption{HT contribution to charged pion pair production
$p\bar {p}\to\pi^{+}\pi^{-}X$ cross section $\Sigma_{\pi^+\pi^-}^{HT}$ as
 a function of the transverse momentum $p_{T}$ for momentum cut-off parameter $\triangle p=0.3$ GeV/\emph{c}, at $\sqrt s$=20 GeV and $y=0$} \label{fig:fig6}
\end{figure}

\begin{figure}[!hbt]
\includegraphics[width=8.6 cm]{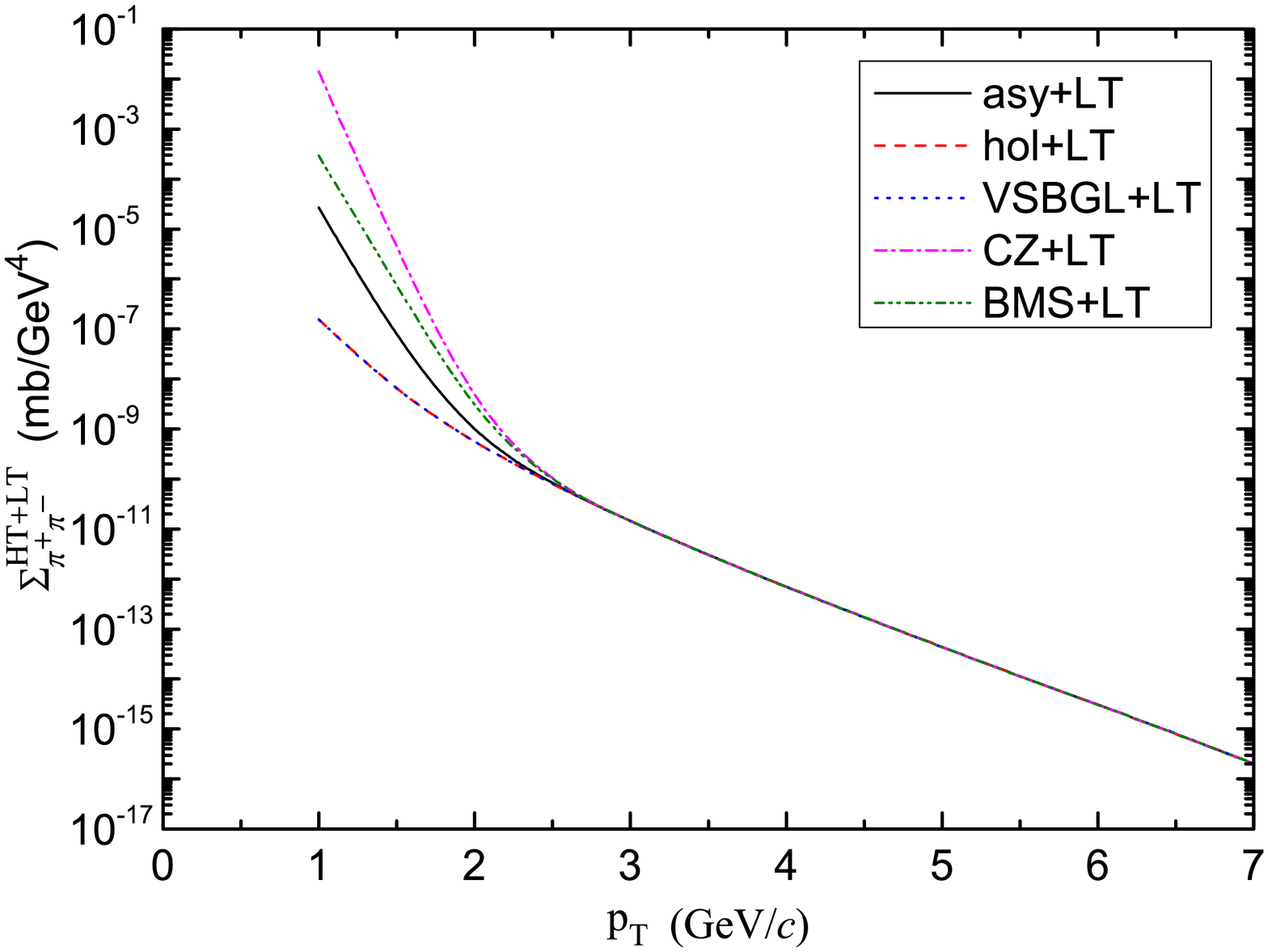}
\caption{The sum of HT and LT  contribution to charged pion pair
production $p\bar {p}\to\pi^{+}\pi^{-}X$  cross section
$\Sigma_{\pi^+\pi^-}^{HT+LT}$ as a function of the transverse
momentum $p_{T}$ for momentum cut-off parameter $\triangle
p=0.3GeV/c$, at $\sqrt s$=20 GeV and $y=0$. Notice that curves for
asy, hol, VSBGL, CZ and BMS pion distribution amplitudes in the
region $2.5\,\,$GeV/\emph{c}$<p_T<7\,\,$GeV/\emph{c} completely
overlap.} \label{fig:fig7}
\end{figure}

\begin{figure}[!hbt]
\includegraphics[width=8.6 cm]{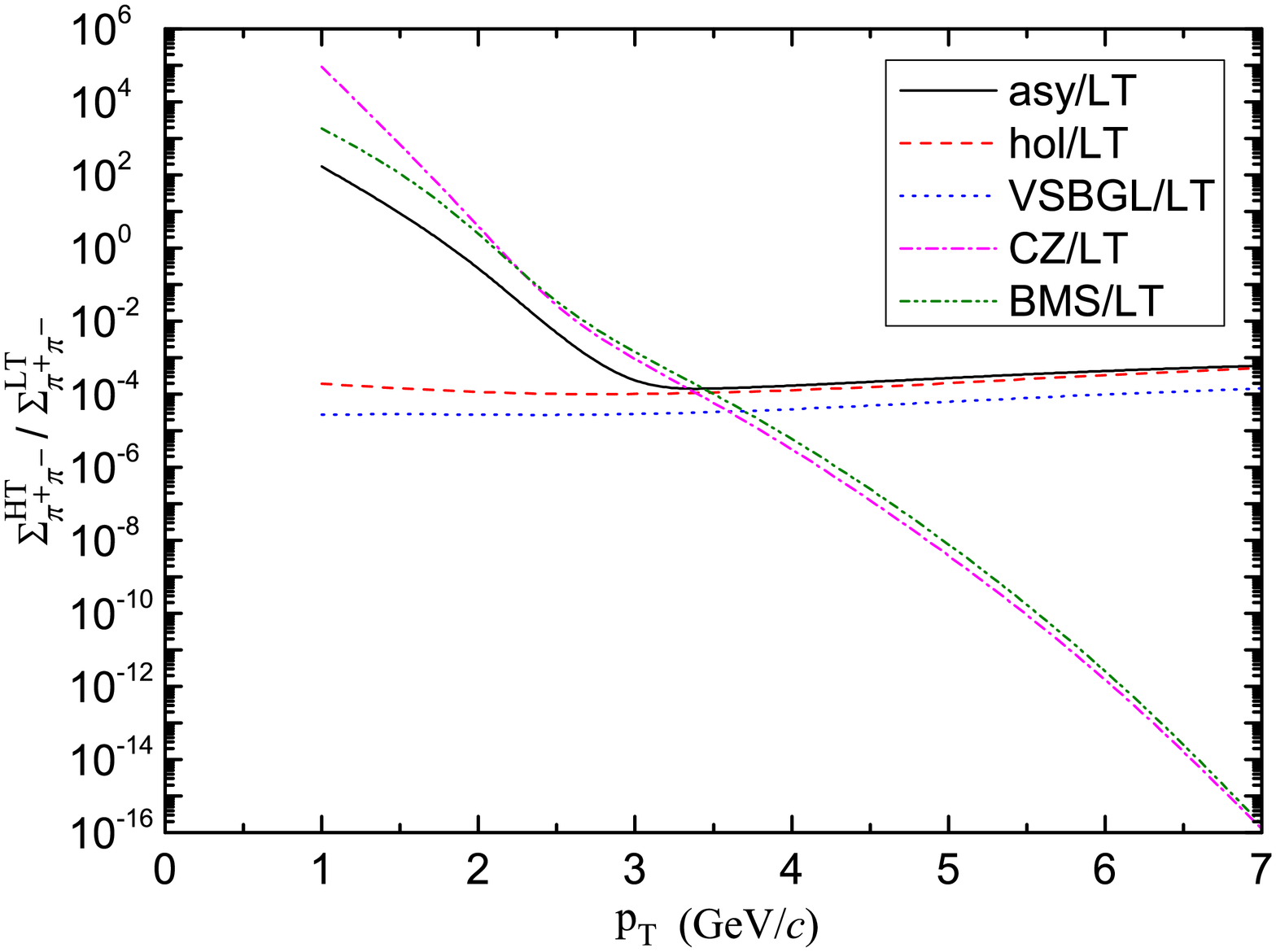}
\caption{Ratio $\Sigma_{\pi^+\pi^-}^{HT}/\Sigma_{\pi^+\pi^-}^{LT}$
as a function of the transverse momentum $p_{T}$ of the pion pair at
the $<{q_{T}^2}>$=0.25GeV$^2$/\emph{c}$^2$, at the c.m. energy
$\sqrt s$=20\,\, GeV and $y=0$} \label{fig:fig8}
\end{figure}

\begin{figure}[!hbt]
\includegraphics[width=8.6 cm]{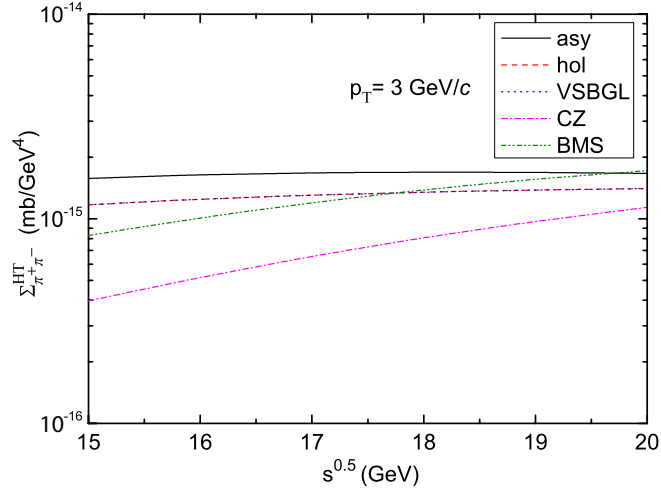}
\caption{HT contribution to charged pion pair production $p\bar
{p}\to\pi^{+}\pi^{-}X$  cross section $\Sigma_{\pi^+\pi^-}^{HT}$ as
a function of the center-of-mass energy $\sqrt s$ at the
$<{q_{T}^2}>$=0.25GeV$^2$/\emph{c}$^2$ and $y=0$.} \label{fig:fig9}
\end{figure}

\begin{figure}[!hbt]
\includegraphics[width=8.6 cm]{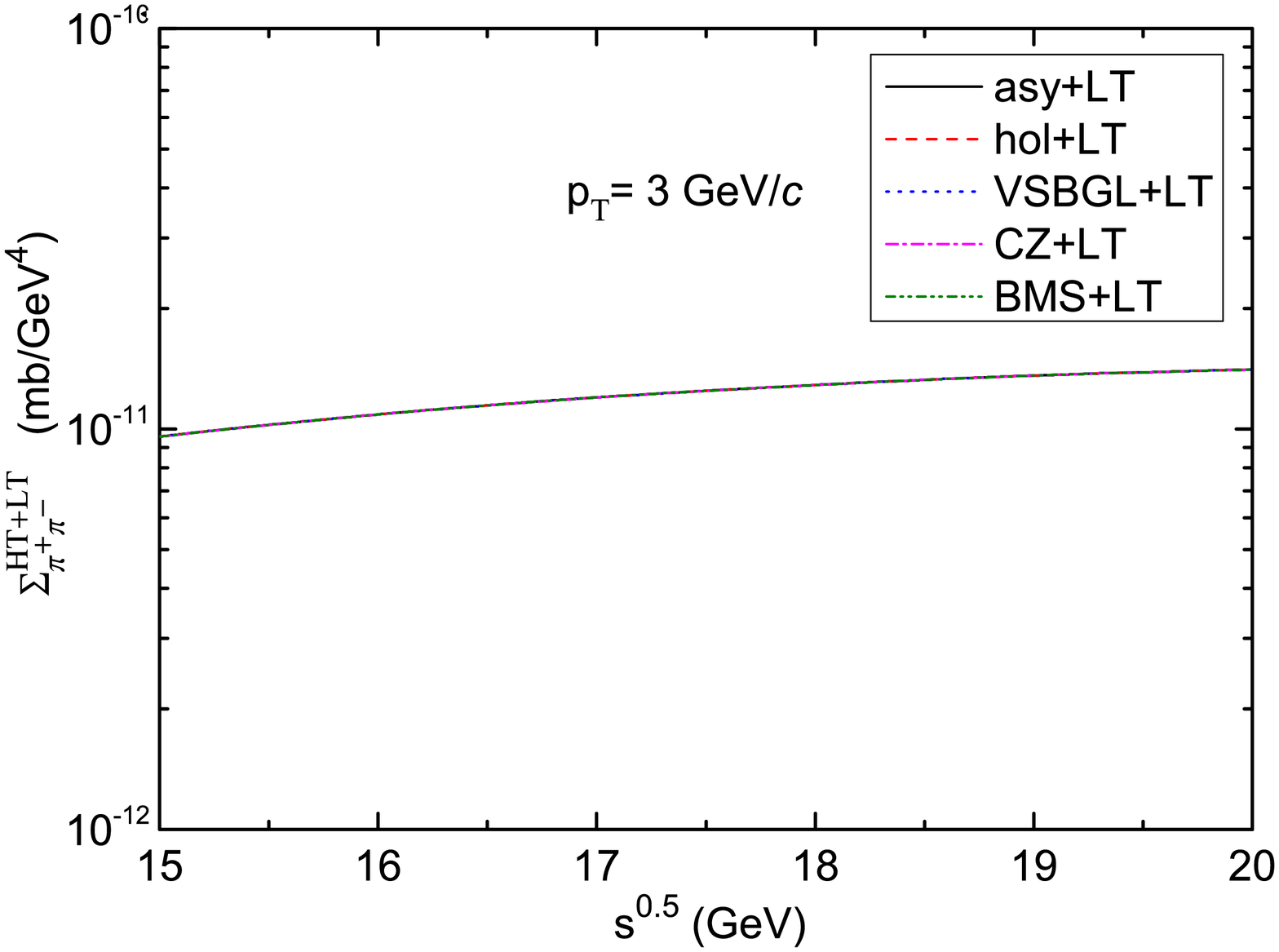}
\caption{The sum of HT and LT  contribution to charged pion pair
production $p\bar {p}\to\pi^{+}\pi^{-}X$  cross section
$\Sigma_{\pi^+\pi^-}^{HT+LT}$ as a function of the center-of-mass
energy $\sqrt s$ for momentum cut-off parameter $\triangle
p=0.3$GeV/\emph{c} and $y=0$. Notice that curves for asy, hol,
VSBGL, CZ and BMS pion distribution amplitudes completely overlap.}
\label{fig:fig10}
\end{figure}

\begin{figure}[!hbt]
\includegraphics[width=8.6 cm]{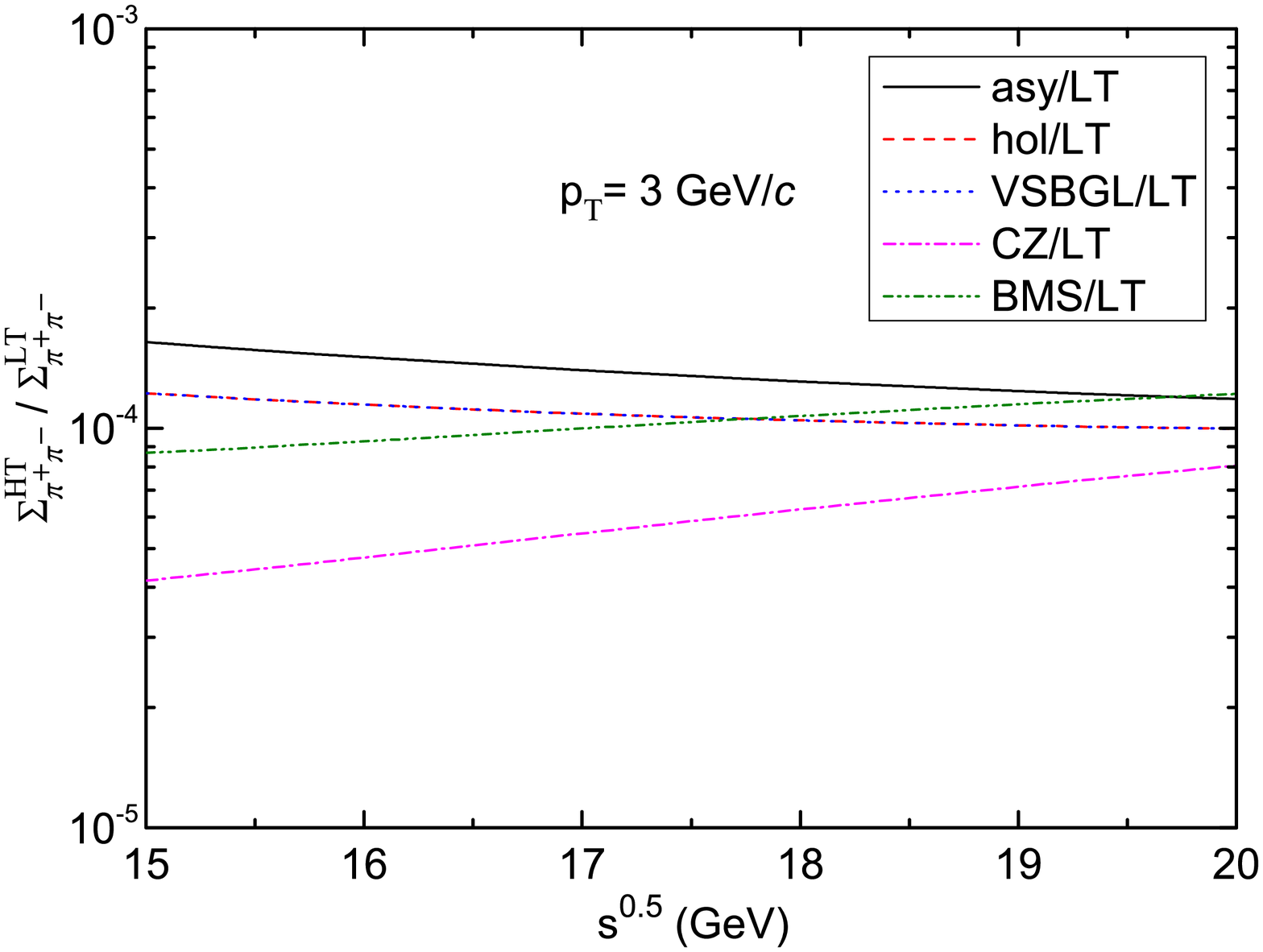}
\caption{Ratio $\Sigma_{\pi^+\pi^-}^{HT}/\Sigma_{\pi^+\pi^-}^{LT}$
as a function of the center-of-mass energy $\sqrt s$ at the
$<{q_{T}^2}>$=0.25GeV$^2$/\emph{c}$^2$ and $y=0$.} \label{fig:fig11}
\end{figure}

\begin{figure}[!hbt]
\includegraphics[width=8.6 cm]{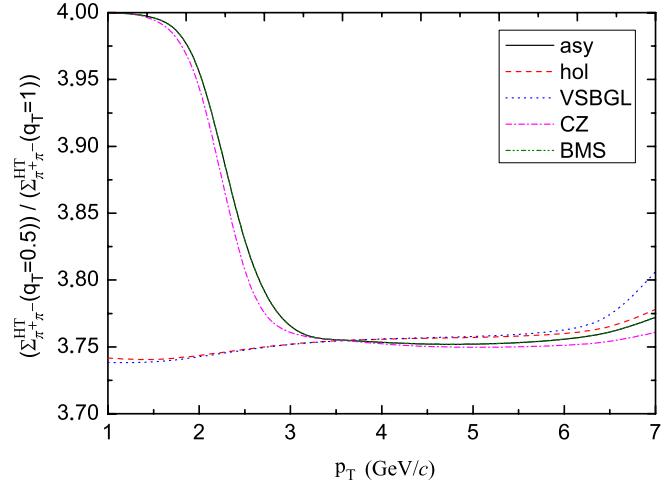}
\caption{Ratio HT cross sections $\Sigma_{\pi^+\pi^-}^{HT}$ is
calculated  with $<{q_{T}^2}>$=0.25 GeV$^2/c^2$ and $<{q_{T}^2}>$= 1
GeV$^2/c^2$ as a function of the pion pair transverse momentum
$p_{T}$ at $\sqrt s$=15 GeV and $y=0$.} \label{fig:fig12}
\end{figure}

\begin{figure}[!hbt]
\includegraphics[width=8.6 cm]{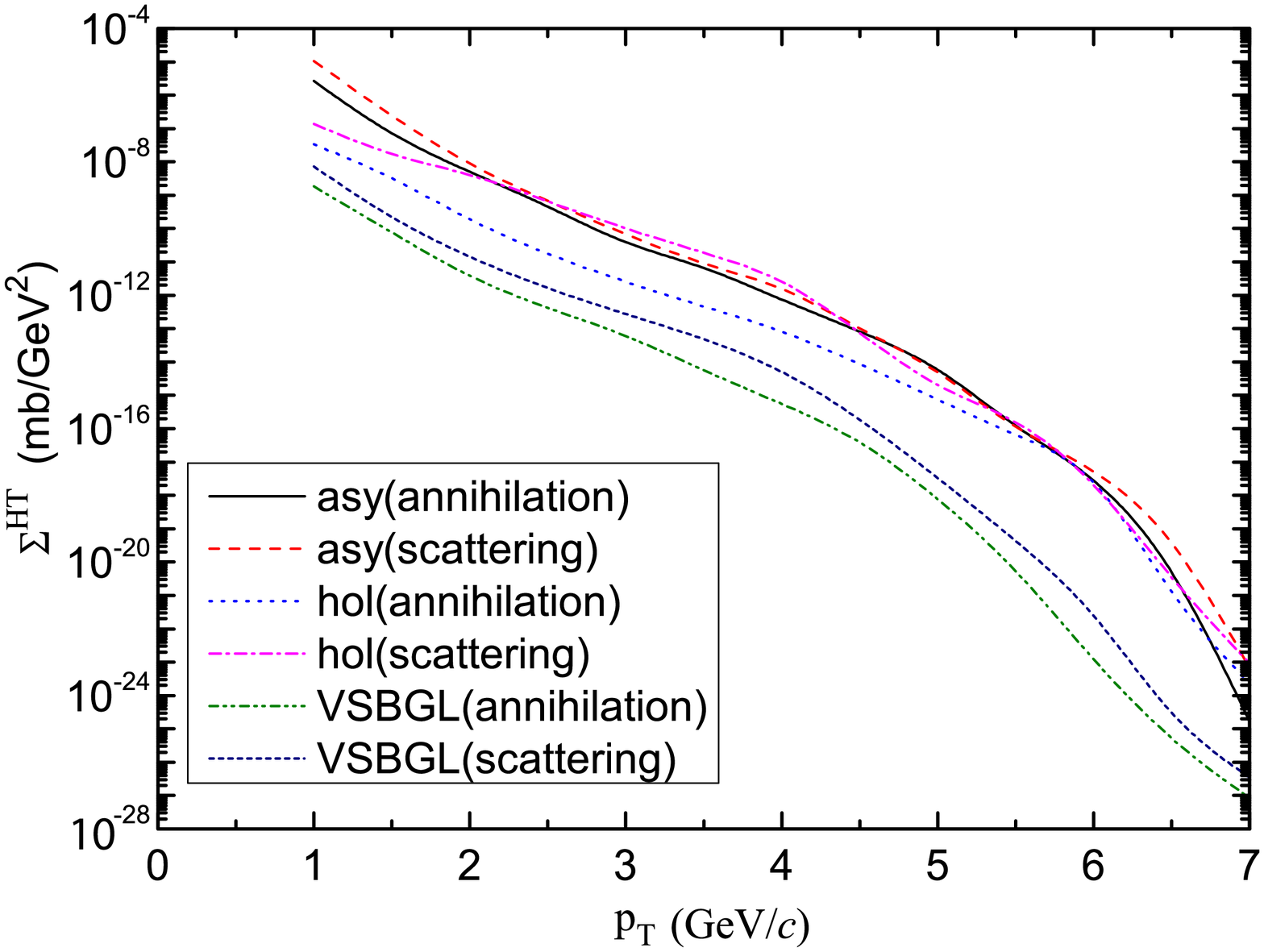}
\caption{HT $p\bar {p}\to\pi^{+}\pi^{-}$ pion pair production  and
$\pi p \to\pi p$ cross sections  as a function of the transverse
momentum $p_{T}$ of the pion  for $<{q_{T}^2}>$=0.25GeV$^2/c^2$,  at
the c.m. energy $\sqrt s$=15  GeV and $y=0$.} \label{fig:fig13}
\end{figure}

\begin{figure}[!hbt]
\includegraphics[width=8.6 cm]{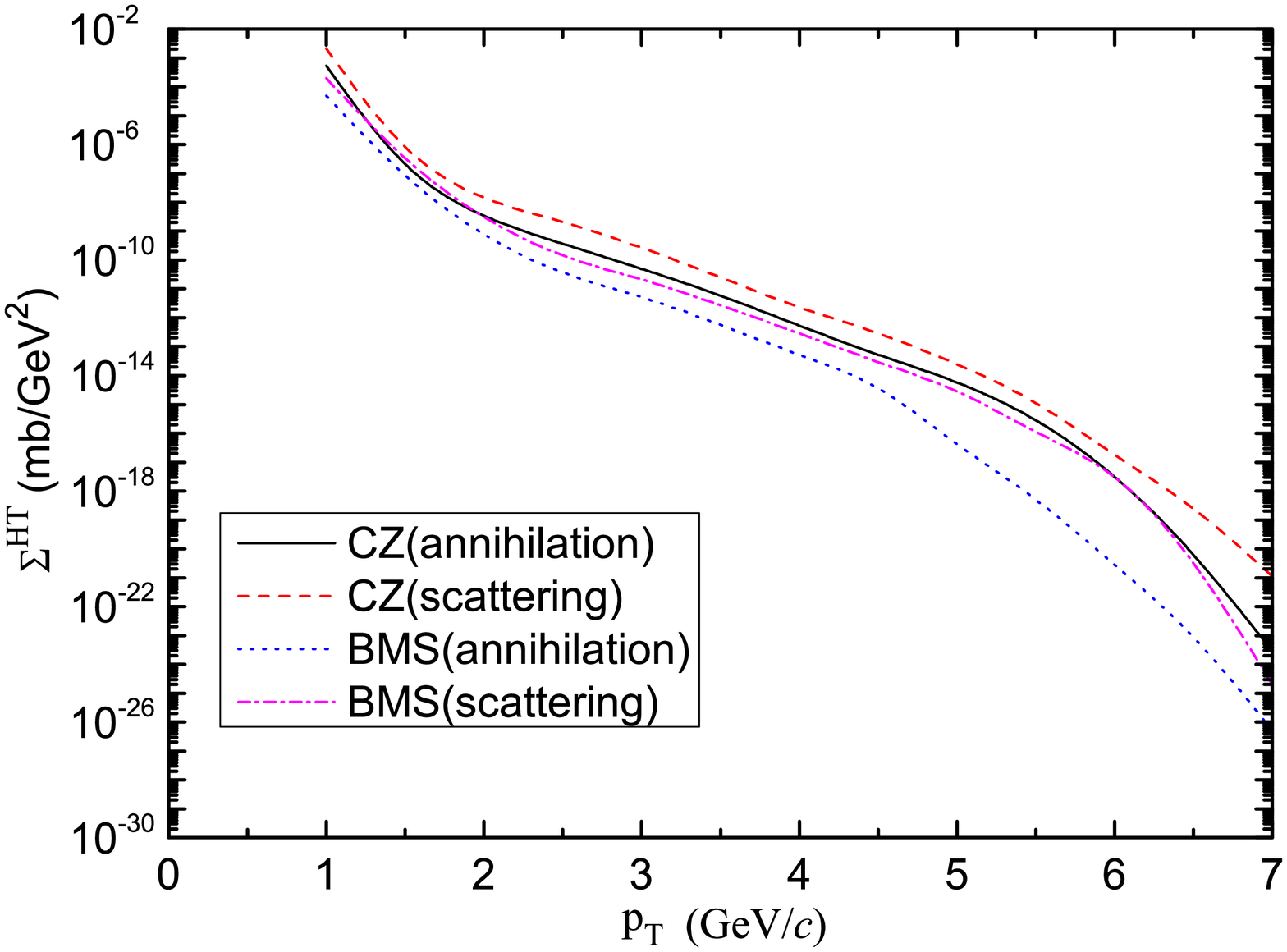}
\caption{HT $p\bar {p}\to\pi^{+}\pi^{-}$ pion pair production  and
$\pi p \to\pi p$ cross sections  as a function of the transverse
momentum $p_{T}$ of the pion for $<{q_{T}^2}>$=0.25GeV$^2/c^2$, at
the c.m. energy $\sqrt s$=15 GeV and $y=0$.} \label{fig:fig14}
\end{figure}

\begin{figure}[!hbt]
\includegraphics[width=8.6 cm]{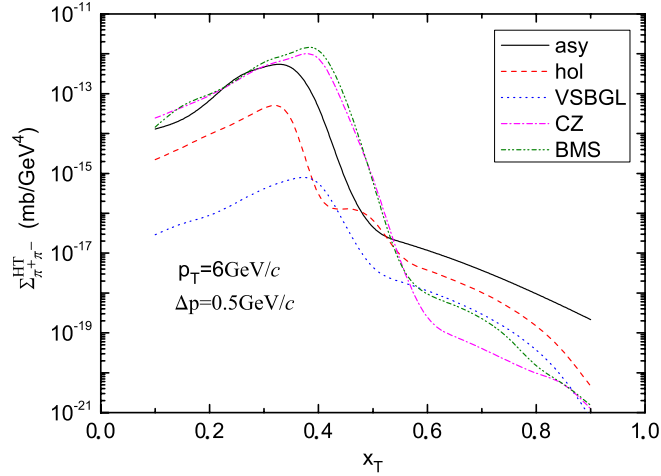}
\caption{HT $p\bar {p}\to\pi^{+}\pi^{-}$ pion pair production  cross
section  as a function of the variable  $x_{T}$ for momentum cut-off
parameter  $\Delta p=0.5$GeV/$c$ at $p_T=6$GeV/$c$ and $y=0$.}
\label{fig:fig15}
\end{figure}

\begin{figure}[!hbt]
\includegraphics[width=8.6 cm]{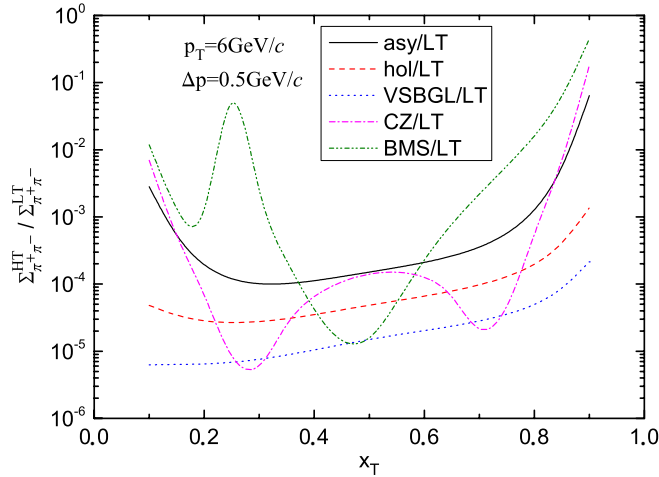}
\caption{Ratio of HT to LT contributions as a function of the
variable  $x_{T}$ for momentum cut-off parameter  $\Delta
p=0.5$GeV/$c$ at $p_T=6$GeV/$c$ and $y=0$.} \label{fig:fig16}
\end{figure}

Let us now discuss in detail the numerical predictions of the HT and
LT cross sections of the pion pair production process
$p\bar{p}\to\pi^{+}\pi^{-}X$ at the PANDA energies taking into
account the full leading-order contributions from quark-antiquark
annihilation and gluon-gluon fusion. We denote the HT cross section
by $\Sigma_{\pi^{+} \pi^{-}}^{HT}$, the LT cross section by
$\Sigma_{\pi^{+} \pi^{-}}^{LT}$, and the sum of HT and LT by
$\Sigma_{\pi^{+} \pi^{-}}^{HT+LT}$. For the quark and gluon
distribution functions inside the proton and antiproton, the
MSTW2008 PDFs \cite{watt} and the quark and gluon fragmentation
functions \cite{Kniehl}  are used. Also, the following abbreviations
are defined: asy is $\Phi_{asy}(x)$, hol is $\Phi^{hol}(x)$, VSBGL
is $\Phi_{VSBGL}^{hol}(x)$, CZ is $\Phi_{CZ}(x,Q^2)$, and BMS is
$\Phi_{BMS}(x,Q^2)$. The results are given for $\sqrt s$ = 15 and 20
GeV  on the transverse momentum $p_T$ ranging from 1 GeV/\emph{c} to
7 GeV/\emph{c} which are also valid for the PANDA experiment.
Obtained results are visualized through in Figs. \ref{fig:fig3} -
 \ref{fig:fig16}.

Firstly, we compare the HT and LT cross sections obtained within
holographic QCD and pQCD. In Figs. \ref{fig:fig3} and 4, we show the
HT cross section $\Sigma_{\pi^+\pi^-}^{HT}$ and the sum of HT and LT
cross sections $\Sigma_{\pi^+\pi^-}^{HT+LT}$ which are calculated in
the context of the FCC approach as a function of the pion pair
transverse momentum $p_{T}$ for the pion DAs for Eqs. (\ref{asy}) -
(\ref{BMS}) and for $y=0$. It is also seen that the
$\Sigma_{\pi^+\pi^-}^{HT}$ and $\Sigma_{\pi^+\pi^-}^{HT+LT}$ cross
sections are monotonically decreasing with an increase in the
transverse momentum of the pion pair. It is worth to mention that at
the c.m. energy $\sqrt s$=15\,\, GeV the maximum value of the frozen
cross section of the process $p\bar{p}\to\pi^{+}\pi^{-}X$ for the
$\Phi_{CZ}(x,Q^2)$ decreases from the interval
$2.10992\times10^{-3}$ mb/GeV$^{4}$ to $1.32239\times10^{-33}$
mb/GeV$^{4}$, but the $\Sigma_{\pi^+\pi^-}^{HT+LT}$ cross sections
for the same DA decreases from  $2.11018\times10^{-3}$  mb/GeV$^{4}$
to $2.26384\times10^{-20}$ mb/GeV$^{4}$. From these results one can
observes that HT cross section  of the pion pair production in the
proton-antiproton collisions appears in the range and should be
observable at the PANDA experiment.

In Fig. \ref{fig:fig5}, we show the ratio
$\Sigma_{\pi^+\pi^-}^{HT}/\Sigma_{\pi^+\pi^-}^{LT}$  for the process
$p\bar{p}\to\pi^{+}\pi^{-}X$ as a function of $p_{T}$ for the pion
DAs given in Eqs. (\ref{asy}) - (\ref{BMS}) at $y=0$. It is seen
that in the region $1\,\,$GeV/\emph{c}$<p_T<3\,\,$GeV/\emph{c}, the
ratio $\Sigma_{\pi^+\pi^-}^{HT}/\Sigma_{\pi^+\pi^-}^{LT}$ for
$\Phi^{hol}(x)$ is enhanced by about one order of magnitude relative
to the $\Phi_{VSBGL}(x)$. However, the enhancement are half an order
of magnitude for $\Phi_{BMS}(x,Q^2)$ and $\Phi_{CZ}(x,Q^2)$, but in
the region $3\,\,$GeV/\emph{c}$<p_T<7\,\,$GeV/\emph{c} the magnitude
relative for $\Phi^{hol}(x)$ and $\Phi_{asy}(x)$ pion distribution
amplitudes is equal.



Through Fig \ref{fig:fig6} - \ref{fig:fig8}, we have displayed the
$\Sigma_{\pi^+\pi^-}^{HT}$, and $\Sigma_{\pi^+\pi^-}^{HT+LT}$ cross
sections and the ratio
$\Sigma_{\pi^+\pi^-}^{HT}/\Sigma_{\pi^+\pi^-}^{LT}$ which are
calculated in the context of the FCC approach as a function of the
pion pair transverse momentum $p_{T}$ for the pion DAs for Eqs.
(\ref{asy}) - (\ref{BMS}), and again for $y=0$ and at the
center-of-mass energy $\sqrt s$= 20 GeV. It is seen from Figs.
\ref{fig:fig6} and  \ref{fig:fig8} that the
$\Sigma_{\pi^+\pi^-}^{HT}$, and $\Sigma_{\pi^+\pi^-}^{HT+LT}$ cross
sections are monotonically decreasing with an increase in the
transverse momentum of the pion pair. In the region
$1\,\,$GeV/\emph{c}$<p_T<7$GeV/\emph{c}, the frozen cross section of
the process $p\bar{p}\to\pi^{+}\pi^{-}X$ decreases from
$1.41213\times10^{-2}$ mb/GeV$^{4}$ to $1.107\times10^{-19}$
mb/GeV$^{4}$, but the sum of HT and LT cross section decreases from
$1.41214\times10^{-2}$ mb/GeV$^{4}$ to $2.01712\times10^{-16}$
mb/GeV$^{4}$.

For the region $1\,\,$GeV/$c<p_T<4\,\,$GeV$/c$, the LT cross section
is enhanced by about four orders of magnitude relative to the HT
cross section calculated in the FCC approach. However, the
$4\,\,$GeV/$c<p_T<7\,\,$GeV/$c$ region with increasing transverse
momentum of the pair pion cross section increases, and the
difference between leading and HT cross sections decreases
essentially. Through Figs. \ref{fig:fig9} - \ref{fig:fig11}, the
dependence of the $\Sigma_{\pi^+\pi^-}^{HT}$ and
$\Sigma_{\pi^+\pi^-}^{HT+LT}$ cross sections and the ratio
$\Sigma_{\pi^+\pi^-}^{HT}/\Sigma_{\pi^+\pi^-}^{LT}$ of the
center-of-mass energy $\sqrt s$ for the pion DAs are displayed by
using  Eqs. (\ref{asy}) - (\ref{BMS}) at $y=0$. Hereby, these
figures indicate that the HT, sum of HT and LT cross sections, and
the ratio increase slowly and smoothly when increasing the beam
energy from 15 GeV to 20 GeV for each pion DAs. In Fig
\ref{fig:fig12} we show that the ratio HT cross section
$\Sigma_{\pi^+\pi^-}^{HT}$ is calculated with $<{q_{T}^2}>$=0.25
GeV$^2/c^2$ and $<{q_{T}^2}>$= 1 GeV$^2/c^2$ as a function of the
pion pair transverse momentum $p_{T}$ for the pion DAs for Eqs.
(\ref{asy}) - (\ref{BMS}), at $y=0$ and the center-of-mass energy
$\sqrt s= 15$ GeV.

One can also observe that the HT cross section in the region
$1\,\,$GeV/$c<p_T<3\,\,$GeV/$c$ decreases more quickly for the DAs
of asy, CZ, BMS with increasing $p_T$, but in the region
$1\,\,$GeV/$c<p_T<7\,\,$GeV/$c$ increases more slowly and smoothly
for the DAs hol, VSBGL with increasing $p_T$. In Figs.
\ref{fig:fig13} and  \ref{fig:fig14}, the comparison of the HT cross
section $\Sigma^{HT}$ is displayed for the proton-antiproton
annihilations into charged pion pairs $p\bar{p}\to\pi^{+}\pi^{-}$
and elastic scattering $\pi p \to\pi p$ processes which are
calculated in the context of the FCC approach as a function of the
pion pair transverse momentum $p_{T}$ for the pion DAs at $y=0$ and
the center-of-mass energy $\sqrt s$= 15 GeV. We can see from Figs.
\ref{fig:fig13} and \ref{fig:fig14} that the HT cross section of the
elastic scattering $\pi p \to\pi p$ process is enhanced by about
half an order of magnitude relative to the
$p\bar{p}\to\pi^{+}\pi^{-}$ cross sections for all pion DAs.


In Figs. \ref{fig:fig15} and \ref{fig:fig16}, we have displayed the
HT and ratio HT to LT cross sections with the dependence on the
variable $x_T$ ranging from $10^{-1}$ to $0.9$ at the $p_T=6$
GeV/$c$ with rapidities of pions $y_1=y_2=0$ for momentum cut-off
parameter $\Delta p=0.5$ GeV/$c$. As is seen from Fig.
\ref{fig:fig15}, the HT cross section in the region
$0.1\,\,<x_T<0.4$ is monotonically increasing with an increase in
the variable  $x_T$. Approximately, the HT cross section for all DAs
has a maximum at the point $x_T=0.4$. After this, the HT cross
section with increasing $x_T$ is decreasing. But, the ratio of HT to
LT cross sections for the dependence on the variable $x_T$ has a
different distinctive behavior. As is seen from Fig \ref{fig:fig16},
the ratio for the $\Phi_{CZ}(x,Q^2)$ and $\Phi_{BMS}(x,Q^2)$  has
two minima and one maximum. The analysis of our calculations shows
that the main reason for this depends on the phenomenological
factors. These plots reveal that the distribution of variable $x_T$
also demonstrates the same dominant contributions in view of DAs as
the ones in the transverse momentum dependence of the cross section.
The ratio of HT to LT contributions remains almost nonstable in a
large interval of $x_T$. This means that the ratio is more sensitive
according to varying the $x_T$. Analysis of our calculations shows
that the HT cross section $\Sigma_{\pi^+\pi^-}^{HT}$  and the ratio
$\Sigma_{\pi^+\pi^-}^{HT}/\Sigma_{\pi^+\pi^-}^{LT}$ are sensitive to
pion DA as predicted in the holographic and pQCD.

\section{CONCLUSIONS}\label{conc}

In this study, the HT contributions, which are included in the
direct and semi-direct productions of the hard scattering process,
to large-$p_T$ pion pair production in proton-antiproton collisions
were discussed in detail. Furthermore, the dependence of HT
contributions on pion-DAs predicted by the light-cone formalism and
the light-front holographic AdS/CFT approach was addressed as well.
It can be also concluded that the results which significantly depend
on the DAs of the pion can be used for their research. The basic
size of the HT cross sections was different depending on the choice
of DAs of the produced pions and also some other phenomenological
factors. Also, for the region $1\,\,$GeV/$c<p_T<3\,\,$GeV/$c$ DAs of
CZ, BMS, in the region $3\,\,$GeV$/c<p_T<7\,\,$GeV/$c$  hol, VSBGL
gave the result which is close in shape to those for the asymptotic
DA, but the HT contributions for CZ were larger than them by one
order of magnitude relative of the asy and 2 - 3 orders for other
DA. However, the ratio of HT to LT contributions allowed us to
determine these regions in the phase space where HT contributions
are essentially observable. This ratio is sensitive to the
transverse momentum $p_T$ and the momentum cut-off parameter
$\triangle p$, which is the detection limit for accompanying
particles. For a small value of  $p_T$, HT contributions yield the
considerably high values. Its effect became significant at the small
$p_T$ region compared to the LT contribution. It should be noted
that semi-direct pion pair production and double jet fragmentation
to pion pair cross section strongly depend on the fragmentation
function of the quark and gluon to pion. Also, the production of
hadrons with large transverse momentum was dominated by the
fragmentation of partons which is produced in parton-parton
scattering with large momentum. The production cross section for
this hard scattering depends on the initial distribution of partons
in the colliding species, the elementary parton-parton cross section
and the fragmentation process of partons into hadrons.

The HT cross section obtained in our study should be observable at a
hadron collider. Also, the feature of HT effects can help
theoretical interpretations of the future PANDA experimental data
for the direct inclusive pion pair production cross section in the
proton-antiproton collisions. As a result, it can be indicated that
the HT processes for large-$p_T$ pion pair production have a key
enabling contribution, where the pions are generated directly in the
hard-scattering subprocess, rather than by gluon and quark
fragmentation. Inclusive pion pair production provides an essential
test case where HT contributions dominate those of LT in the certain
kinematic regions. The HT contributions can be utilized to interpret
theoretically the future experimental data for the charged pion pair
production in $p\bar p$ collisions. The results of this work can be
useful to provide a simple test of the short distance structure of
QCD as well as to determine more precise DAs of the pion.

\section*{\bf Acknowledgments}
A. I. A. is grateful for the financial support by the Science
Development Foundation under the President of the Republic of
Azerbaijan - \textbf{Grant
No.EIF/MQM/Elm-Tehsil-1-2016-1(26)-71/11/1} and  Baku State
University \textbf{Grant No. "50+50" (2018 - 2019)}. The authors are
grateful to S. V. Mikhailov for useful discussions.

\newpage


\end{document}